\documentclass[showpacs,floatfix,amssymb,amsmath,aps,prc,twocolumn,eqsecnum,nofootinbib]{revtex4}
\usepackage[dvips]{graphicx}
\usepackage{latexsym,epsfig,bm,times,psfrag,subfigure}
\usepackage{dcolumn}
\usepackage{color}
\renewcommand\a{\alpha}
\renewcommand\b{\beta}
\renewcommand\d{\delta}

\renewcommand\l{\lambda}
\renewcommand\r{\rho}

\renewcommand\t{\tau}

\renewcommand\c{\chi}
\renewcommand\j{\psi}
\renewcommand\o{\omega}
\newcommand\e{\epsilon}
\newcommand\g{\gamma}
\newcommand\z{\zeta}
\newcommand\m{\mu}
\newcommand\n{\nu}
\newcommand\x{\xi}
\newcommand\p{\pi}

\newcommand\s{\sigma}

\newcommand\w{\eta}
\newcommand\ve{\varepsilon}

\renewcommand\S{\Sigma}
\renewcommand\O{\Omega}

\newcommand\D{\Delta}

\newcommand\J{\Psi}

\newcommand{\fig}[1]{Fig.~\ref{#1}}
\newcommand{\eq}[1]{Eq.~(\ref{#1})}

\newcommand\lb{\left(}
\newcommand\rb{\right)}
\newcommand\ls{\left[}
\newcommand\rs{\right]}

\newcommand{\lan}{\langle}
\newcommand{\ran}{\rangle}
\newcommand\ua{\uparrow}
\newcommand\da{\downarrow}
\newcommand\ra{\rightarrow}

\newcommand{\non}{\nonumber\\}
\newcommand\pt{\partial}
\newcommand{\idp}[2]{\int\frac{d^{\,#1}#2}{(2\p)^#1}}
\newcommand{\ie}{\emph{i.e.}}

\newcommand{\eg}{\emph{e.g.}}
\newcommand{\etc}{\emph{etc}}
\newcommand{\etal}{\emph{et al.}}
\newcommand{\cl}{{\cal L}}

\newcommand{\mf}{{\rm mf}}

\newcommand{\bx}{{\mathbf x}}
\newcommand{\br}{{\mathbf r}}

\newcommand{\bk}{{\mathbf k}}
\newcommand{\bq}{{\mathbf q}}
\renewcommand{\sf}{{\rm{sf}}}

\newcommand{\bcs}{{\rm{BCS}}}
\newcommand{\pg}{{\rm{pg}}}
\newcommand{\cs}{{\cal S}}
\newcommand{\cf}{{\cal F}}
\newcommand{\cg}{{\cal G}}
\newcommand{\jb}{{\bar \j}}
\newcommand{\Jb}{{\bar \J}}
\newcommand{\db}{{\bar \D}}

\begin{document}

\title{BCS-BEC Crossover in Symmetric Nuclear Matter at Finite Temperature: Pairing Fluctuation and Pseudogap}
\author{\normalsize{Xu-Guang Huang$^{1,2}$}}
\affiliation{$^1$ Frankfurt Institute for Advanced Studies, D-60438 Frankfurt am Main, Germany\\
$^2$ Institut f\"ur Theoretische Physik,
Goethe-Universit\"at,D-60438 Frankfurt am Main, Germany}

\date{\today}

\begin{abstract}
By adopting a $T$-matrix-based method within the $G_0G$
approximation for the pair susceptibility, we studied the effects of
pairing fluctuation on the BCS-BEC crossover in symmetric nuclear
matter. The pairing fluctuation induces a pseudogap in the
excitation spectrum of nucleon in both superfluid and normal phases.
The critical temperature of the superfluid transition was
calculated. It differs from the BCS result remarkably when density
is low. We also computed the specific heat, which shows a nearly
ideal BEC-type temperature dependence at low density, but a BCS-type
behavior at high density. This qualitative change of the temperature
dependence of specific heat may serve as a thermodynamic signal for
the BCS-BEC crossover.
\end{abstract}
\pacs{21.65.-f, 21.30.Fe, 26.60.-c, 74.20.-z}

\maketitle

\section {Introduction}\label{introduction}
One of the most common properties of attractive fermion many-body
system is the arising of superfluid state at low temperature.
Depending on the strength of the attractive interaction between two
fermions, however, the physical contents of the superfluid state
could be distinguishably altered. When the interaction is weak, the
system can be well described by the Bardeen-Cooper-Schrieffer (BCS)
theory. In this case, the superfluidity is due to the condensate of
loosely correlated Cooper pairs and superfluid gap is much smaller
than the Fermi energy. When the interaction becomes sufficiently
strong, the two-fermion bound state could form, which may behave
like a boson. In this situation, the superfluidity is due to the
Bose-Einstein condensation (BEC) of the tightly bound two-fermion
state and the superfluid gap could be much larger than the Fermi
energy. Although the BCS and BEC limits have quite different
physics, it was found that there is no true phase transition
(traditional symmetry breaking) happening in between. The transition
from BCS state to BEC state is smooth and is often called the
BCS-BEC
crossover~\cite{Eagles:1969zz,Leggett:1980,Nozieres:1985zz,Melo:1993}.
Such BCS-BEC crossover was recently realized in cold atomic
experiments (see Refs.~\cite{Giorgini:2008zz,Bloch:2008zzb} and
references therein).

It was well known that cold nuclear matter can be in a superfluid
state, which plays a crucial role in a variety of nuclear many-body
problems, from neutron stars, low-energy heavy-ion collisions, to
finite nuclei. It was argued that the BCS-BEC crossover should also
occur in nuclear matter where the BCS state of neutron-proton ($np$)
Cooper pairs at high density undergoes a smooth transition into BEC
state of deuterons at low
density~\cite{Alm:1993zz,Stein:1994,Baldo:1995zz,Lombardo:2001ek,Isayev:2004aw,Sedrakian:2005db,Isayev:2006qb,Mao:2008wz}.
At the same time, the chemical potential changes its sign at a
certain density and finally approaches one-half of the deuteron
binding energy at the low-density limit. Recently, a similar
situation was also studied for the neutron-neutron ($nn$) pairs in
the $^1S_0$
channel~\cite{Hansen:1987mc,Zhukov:1993aw,Matsuo:2004pr,Matsuo:2005vf,Hagino:2005we,Hagino:2006ib,Margueron:2007uk,Isayev:2008wu,Hagino:2008vm,Mao:2008wz,Sun:2009iw,Wlazlowski:2009yi}.
It was found that in certain (low) density regions the $nn$ pairs
can be strongly correlated. However, no assured BEC state was found
for $nn$ pairs.

So far, most of the investigations of nuclear BCS-BEC crossover in
the literatures focused on the ground state crossover described by
BCS theory. Although the BCS theory succeeds in describing the
BCS-BEC crossover at zero temperature, it, as a mean-field theory,
is not sufficient to describe low-density nuclear matter at finite
temperature where the pairing fluctuation is substantial due to the
strongly correlating nature of the system. Actually, as a
consequence of the strong correlation, the low-density nuclear
matter exhibits ``pseudogap" phenomena above the critical
temperature $T_c$ of superfluid transition and has an exotic normal
state that is different from the Fermi liquid normal state
associated with BCS
theory~\cite{Bozek:1999rv,Schnell:1999tu,Bozek:2000fn}. Similar
situations were also found in other strongly correlated systems,
such as high $T_c$
superconductors~\cite{Ding:1996,Emery:1995,Chakravarty,Lee,Chen:1998zz,Loktev:2000ju}
and cold atomic Fermi gases under Feshbach
resonances~\cite{PhysRevB58.R5936,phyc1999,prb61:11662,Loktev:2000ju,phyrept2005,arXiv:0810.1938,Chien:2009}.
To include the pairing fluctuation effects and investigate the
pseudogap phenomena, we will adopt a $T$-matrix formalism based on a
$G_0G$ approximation for the pair susceptibility which was first
introduced by the Chicago group
~\cite{Chen:1998zz,PhysRevB58.R5936,phyc1999,prb61:11662,phyrept2005,arXiv:0810.1938,Chien:2009}.
This formalism generalizes the early works of Kadanoff and
Martin~\cite{Kadanoff} and Patton~\cite{Patton}, and can be
considered as a natural extension of the BCS theory since they share
the same ground state. Moreover, this formalism allows
quasi-analytic calculations and gives a simple physical
interpretation of the pseudogap phase. It clearly shows that the
pseudogap is due to the incoherent pairing fluctuation.

Our focus will be put on the $np$ pair in symmetric nuclear matter
(mainly in the low-density region), since the interaction in this
case is more attractive than in the $nn$ or $pp$ channels and it
provides a very good playground for the BCS-BEC crossover. We will
extend the early
studies~\cite{Alm:1993zz,Stein:1994,Baldo:1995zz,Lombardo:2001ek,Isayev:2004aw,Sedrakian:2005db,Isayev:2006qb,Mao:2008wz}
to include the pairing fluctuation effects and determine the
magnitude of the pseudogap. Furthermore, the transition temperature
for the onset of the superfluid and the thermodynamic properties
will be also be a concern. Such a study will be helpful to
understand the strongly coupling nature of low-density nuclear
matter and may give useful information on the physics of the surface
of nuclei, expanding nuclear matter from heavy-ion collisions,
collapsing stars, \etc.

The article is organized as follows. We give a brief summary of the
effective nucleon-nucleon potential in Sec.~\ref{potential}. In
Sec.~\ref{tmatrix}, we give a detailed theoretical scheme of how the
$T$-matrix-based formalism works at finite temperature. The
numerical results are presented in Sec.~\ref{numerical}. We
summarize our results in Sec.~\ref{summary}. Throughout this
article, we use natural units $\hbar=k_B=c=1$.

\section {Effective Nucleon-Nucleon Potential}\label{potential}
The aim of this article is not to determine the precise values of
the pairing gap, the critical temperature, \etc., but rather to
perform a qualitative (or semi-quantitative) study of the effects of
pairing fluctuation on BCS-BEC crossover. In order to highlight the
essential physics, we will adopt a simple density dependent contact
interaction (DDCI) developed in
Refs.~\cite{Garrido:1999at,Garrido:2000dr}. The potential is of the
form
\begin{eqnarray}
\label{paris1}
V(\bx-\bx')=v_0\left\{1-\w\ls\r(\bx)\over\r_0\rs^{\g}\right\}\d(\bx-\bx'),
\end{eqnarray}
where $v_0, \w, \g$ are three adjustable parameters,
$\r(\bx)=\r_n(\bx)+\r_p(\bx)$ is the nuclear density and $\r_0=0.17
\ \rm{fm}^{-3}$ is the normal nuclear density. Taking suitable
values of the parameters, one can reproduce the pairing gap
$\D(k_F)$ as a function of Fermi momentum $k_F=(3\p^2\r/2)^{1/3}$ in
the channels $L=0,I=1,I_z=\pm1,S=0$ and $L=0,I=0,S=1,S_z=0$
calculated from realistic nucleon-nucleon
potentials~\cite{Garrido:1999at,Garrido:2000dr}, where $L$ is
orbital angular momentum, $I$ denotes isospin, and $S$ is spin.
According to Garrido $\etal$~\cite{Garrido:1999at,Garrido:2000dr},
we will choose in the following numerical calculation $\w=0,
v_0=-530\, \rm{MeV\, fm^3}$ in the $I=0$, $^3S_1$ ($np$ pairing)
channel and a energy cutoff $\e_c=60\, \rm{MeV}$ to regularize the
integration. With these parameters one must use a density-dependent
effective nucleon mass $m(\r)$ corresponding to the Gogny
interaction\cite{Garrido:1999at,Garrido:2000dr},
\begin{eqnarray}
\label{m} \ls m(\r)\over
m_0\rs^{-1}\!\!\!\!\!\!\!&=&\!\!1+\frac{m_0}{2}\frac{k_F}{\sqrt{\p}}\sum_{c=1}^2[W_c+2(B_c-H_c)-4M_c]\non&&\times\m_c^3e^{-x_c}\ls\frac{\cosh{x_c}}{x_c}-\frac{\sinh{x_c}}{x_c^2}\rs,
\end{eqnarray}
where $x_c=k_F^2\m_c^2/2$, $m_0=939\, \rm{MeV}$ is the bare mass of
nucleon, and $\m_c, W_c, B_c, H_c, M_c$ are parameters corresponding
to the Gogny force D1\cite{Ring:1980,Decharge:1979fa}, their values
are listed in Table~\ref{gogny}.
\begin{table}[htb]\renewcommand{\arraystretch}{1.5}\addtolength{\tabcolsep}{4pt}
\caption{Parameters in the effective mass of nucleon (\ref{m})
corresponding to the Gogny interaction
D1\cite{Ring:1980,Decharge:1979fa}.}
\begin{tabular}{cccccc} \hline\hline
c & $\m_c$[fm] & $W_c$[MeV] & $B_c$[MeV] & $H_c$[MeV] & $M_c$[MeV]\\
\hline 1 & 0.7 & $-402.4$ & $-100.0$ & $-496.2$ & $-23.56$\\
2 & 1.2 & $-21.30$ & $-11.77$ & 37.27 & $-68.81$\\
\hline\hline
\end{tabular}\label{gogny}
\end{table}

\section {T-Matrix-Based Formalism}\label{tmatrix}
We consider the nuclear matter as an infinite system of interacting
fermions. In the low-density region, $np$ pairing is realized mainly
in the spin-triplet $s$-wave channel, so let us consider the
following Lagrangian describing neutron and proton interaction via
two-body attractive forces in $^3S_1, S_z=0$ channel,
\begin{eqnarray}
\label{lag}
\cl&=&\sum_{i=,np}\sum_{\s=\ua,\da}\jb_{i,\s}\lb-\pt_\t+\frac{\nabla^2}{2m}+\m\rb\j_{i,\s}\non&&+
\frac{g}{2}\lb\jb_{n\ua}\jb_{p\da}-\jb_{p\ua}\jb_{n\da}\rb\lb\j_{p\da}\j_{n\ua}-\j_{n\da}\j_{p\ua}\rb,\non
\end{eqnarray}
where $g=-v_0>0$ is the coupling strength in $np$ channel and
$\t=it$ is the imaginary time. Introducing auxiliary fields
$\db\equiv (g/2)\lb\jb_{n\ua}\jb_{p\da}-\jb_{p\ua}\jb_{n\da}\rb$ and
$\D\equiv(g/2)\lb\j_{p\da}\j_{n\ua}-\j_{n\da}\j_{p\ua}\rb$, we can
recast \eq{lag} as
\begin{eqnarray}
\label{lag1}
\cl&=&\sum_{i=,np}\sum_{\s=\ua,\da}\jb_{i,\s}\lb-\pt_\t+\frac{\nabla^2}{2m}+\m\rb\j_{i,\s}-\frac{2}{g}\db\D\non&&+
\db\lb\j_{p\da}\j_{n\ua}-\j_{n\da}\j_{p\ua}\rb+\lb\jb_{n\ua}\jb_{p\da}-\jb_{p\ua}\jb_{n\da}\rb\D,\non&=&\Jb
\cs^{-1}\J-\frac{2}{g}\db\D,
\end{eqnarray}
where we have introduced the Nambu-Gorkov spinor
$\J=(\j_{n\ua}\,,\jb_{p\da}\,,\j_{p\ua}\,,\jb_{n\da})^{\rm T}$, and
\begin{eqnarray}
\cs^{-1}\equiv \lb\begin{matrix}\cs^{-1}_1 & 0 \\ 0 & \cs_2^{-1}
\end{matrix}\rb,
\end{eqnarray}
with
\begin{eqnarray}
\cs_1^{-1}\equiv \lb\begin{matrix}
-\pt_\t+\nabla^2/(2m)+\m & \D \\
\db & -\pt_\t-\nabla^2/(2m)-\m
\end{matrix}\rb,\non
\end{eqnarray}
\begin{eqnarray}
\cs_2^{-1}\equiv \lb\begin{matrix}
-\pt_\t+\nabla^2/(2m)+\m & -\D \\
-\db & -\pt_\t-\nabla^2/(2m)-\m
\end{matrix}\rb.\non
\end{eqnarray}
It is seen that $\cs_2$ differs from $\cs_1$ only by minus signs in
front of $\D$ and $\db$. To make our formulae more compact, in the
following discussions we will treat $\cs_1$ only and neglect the
subscript $1$ without confusion.

In the rest of this section, following the works of the Chicago
group, we will introduce the basic method of the $T$ matrix. This
$T$ matrix is defined as an infinite series of ladder-diagrams in
particle-particle channel (rather than particle-hole channel) by
constructing the ladder by one free nucleon propagator and one full
nucleon propagator. Then, as usual, the $T$ matrix enters the
nucleon self-energy in place of the bare interaction vertex. The
coupled $T$-matrix equation and the self-energy equation (as well as
number density equation) should be solved self-consistently. One can
view this approach as the simplest generalization of the BCS scheme,
which formally can also be cast in a $T$-matrix formalism. Let us
discuss this point in the following subsection.
\subsection {BCS theory}\label{bcs}
The BCS theory is based on mean-field approximation to the
(anomalous) self-energy, \ie, $\D$ and $\db$ are chosen as their
mean-field values $\D=\D_\sf$ and $\db=\db_\sf$ (without loss of
generality, we put $\D_\sf$ and $\db_\sf$ to be constants and
$\db_\sf=\D_\sf$), which are regarded as order parameters for
superfluid phase transition. We start with the Nambu-Gorkov
formalism in momentum space,
\begin{eqnarray}
\label{invprop}
\cs_{\mf}^{-1}(K)=\lb\begin{matrix}\cg_0^{-1}(K) & \D_\sf\\
\D_\sf & -\cg_0^{-1}(-K)\end{matrix}\rb,
\end{eqnarray}
where $K=(i\o_n, \bk)$ and $i\o_n=i(2n+1)\p T$ is the fermion
Matsubara frequency. $\cg_0^{-1}(K)=i\o_n-\x_\bk$ is the inverse of
free nucleon propagator, $\x_\bk=\bk^2/(2m)-\m$ is the dispersion
relation of free nucleon. From \eq{invprop} one gets,
\begin{eqnarray}
\label{bcsprop}
\cs_\mf(K)=\lb\begin{matrix}\cg_\mf(K) & \cf_\mf(K)\\
\cf_\mf(K) & -\cg_\mf(-K)\end{matrix}\rb,
\end{eqnarray}
where $\cf_\mf(K)$ is the anomalous propagator,
\begin{eqnarray}
\cf_\mf(K)&=&\D_\sf\cg_\mf(K)\cg_0(-K)\non
&=&\frac{-\D_\sf}{(i\o_n)^2-E_\bk^2},
\end{eqnarray}
and $\cg_\mf(K)$ is the mean-field single nucleon propagator,
\begin{eqnarray}
\cg_\mf(K)&=&\ls\cg_0^{-1}(K)-\S_\mf(K)\rs^{-1}\non
&=&\frac{i\o_n+\x_\bk}{(i\o_n)^2-E_\bk^2}
\end{eqnarray}
with the mean-field dispersion relation of nucleon
$E_\bk=\sqrt{\x_\bk^2+\D^2_\sf}$ and the mean-field self-energy
\begin{eqnarray}
\label{means} \S_\mf(K)=-\D_\sf^2\cg_0(-K).
\end{eqnarray}
The coupled gap and density equations read
\begin{eqnarray}
\D_\sf &=&\frac{g}{\b V}\sum_K\cf_\mf(K)\non
&=&\frac{g\D_\sf}{V}\sum_\bk\frac{1}{2E_\bk}\ls1-2n_F(E_\bk)\rs,\non
\r&=&\frac{2}{\b V}\sum_K e^{i\w\o_n}\cg_\mf(K)\non &=&
\frac{2}{V}\sum_\bk\ls1-\frac{\x_\bk}{E_\bk}\lb1-2 n_F(E_\bk)\rb\rs,
\end{eqnarray}
where $n_F(x)=1/[\exp{(\b x)}+1]$ is the Fermi-Dirac function and
$e^{i\w\o_n}$ with $\w\ra0$ is a convergence factor for Matsubara
summation. The prefactor 2 on the right-hand side of density
equation counts the degeneracy of $\cs_1$ and $\cs_2$.

In BCS theory, $np$ pairs enter into the problem below $T_c$, but
only through their condensates at zero momentum. By rewriting the
mean-field self-energy $\S_\mf(K)$ in a manner of
\begin{eqnarray}
\label{smf} \S_\mf(K)=\frac{1}{\b V}\sum_Q t_\mf(Q)\cg_0(Q-K),
\end{eqnarray}
we are aware of that these {\it condensed} pairs can be associated
with a $T$ matrix in the following form
\begin{eqnarray}
\label{tmf} t_\mf(Q)=-\D^2_\sf\d(Q).
\end{eqnarray}
with $Q=(q_0, \bq), q_0=i\o_\n=i2\n \p T, \n\in \mathbb{Z}$ being
the boson Matsubara frequency and $\d(Q)=\b\d_{\n,0}\d^{(3)}(\bq)$.
Furthermore, if we define the mean-field pair susceptibility as
\begin{eqnarray}
\label{cmf} \c_\mf(Q)=\frac{1}{\b V}\sum_K\cg_\mf(K)\cg_0(Q-K),
\end{eqnarray}
we can re-write the gap equation in superfluid phase as
\begin{eqnarray}
\label{thoulessmf} 1-g\c_\mf(0)=0,\;\; T\leq T_c.
\end{eqnarray}
This suggests that one can consider the {\it uncondensed} pair
propagator or $T$ matrix to be of the form
\begin{eqnarray}
\label{pmf} t_{\rm pair}=\frac{-g}{1-g\c_\mf(Q)},
\end{eqnarray}
and then the gap equation is given by $t_{\rm pair}^{-1}(0)=0$.

It is well known that the critical temperature $T_c$ in the BCS
theory is related to the appearance of a singularity in a $T$ matrix
in the form of \eq{pmf} but with $\D_\sf=0$. This is the so-called
Thouless criterion for $T_c$~\cite{Thouless}. But the meaning of
\eq{thoulessmf} is more general as stressed by Kadanoff and
Martin~\cite{Kadanoff}. It states that under a asymmetric choice of
$\c$, the gap equation is equivalent to the requirement that the $T$
matrix associated with uncondensed pair remains singular at zero
momentum and energy for all temperatures below $T_c$.

Although the construction of the uncondensed pair propagator
(\ref{pmf}) in BCS scheme is quite natural, the uncondensed pair has
no feedback to the nucleon self-energy (\ref{smf}). 
When the coupling is weak, such a feedback is not important, but if
the system is strongly coupled, this feedback will be significant.
The simplest way to include the feedback effects is to replace
$t_\mf$ in \eq{smf} by $t_\mf+t_{\rm pair}$. But to make such an
inclusion self-consistent, $t_{\rm pair}$ should be somewhat
modified which we discuss in next subsection.

\subsection {$G_0G$ formalism at $T\leq T_c$ }\label{g0g}
\begin{figure}[!htb]
\begin{center}
\includegraphics[width=6.5cm]{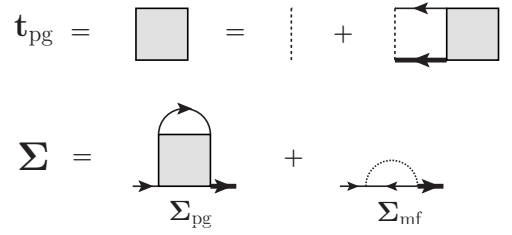}
\caption{Feynman diagrams for the $T$ matrix of non-condensed pairs
and the nucleon self-energy in the $G_0G$ formalism.}
\label{feynman}
\end{center}
\end{figure}

Physically, the BCS theory involves the contribution to nucleon
self-energy below $T_c$ only from those condensed pairs, \ie, the
$\bq=0$ Cooper pairs. This is justified only at weak-coupling
region. Generally, in superfluid phase, the self-energy consists of
two distinctive contributions, one from the superfluid condensate,
and the other from thermal pair excitations. Correspondingly, it is
natural to decompose the self-energy into two additive terms
\begin{eqnarray}
\S(K)=\frac{1}{\b V}\sum_Q t(Q)\cg_0(Q-K)=\S_\mf(K)+\S_\pg(K),\non
\end{eqnarray}
with the $T$ matrix accordingly given by
\begin{eqnarray}
\label{tpg0} t(Q)&=&t_\mf(Q)+t_\pg(Q),\non
t_\pg(Q)&=&\frac{-g}{1-g\c(Q)},
\end{eqnarray}
where the subscript $\pg$ indicates that this term will lead to the
pseudogap in nucleon dispersion relation as will become clear soon.
See \fig{feynman} for the Feynman diagrams for $t_\pg(Q)$ and
$\S(K)$. Comparing with the BCS scheme, $t_\mf(Q)$ in \eq{smf} is
replaced by $t(Q)$, and $\S(K)$ now contains the feedback of
uncondensed pairs. The pair susceptibility $\c(Q)$, as inspired by
\eq{cmf}, is chosen to be the following asymmetric $G_0G$ form,
\begin{eqnarray}
\c(Q)&=&\frac{1}{\b V}\sum_K\cg(K)\cg_0(Q-K).
\end{eqnarray}

In spirit of Kadanoff and Martin, we now propose the superfluid
instability condition or gap equation as [extension of
\eq{thoulessmf}]
\begin{eqnarray}
\label{thouless} 1-g\c(0)=0,\;\; T\leq T_c.
\end{eqnarray}
We stress here that this condition has quite clear physical meaning
in BEC regime. The dispersion relation of the bound pair is given by
$t^{-1}(Q)=0$, hence $t^{-1}(0)\propto\m_b$ with $\m_b$ the
effective chemical potential of the pairs. Then the BEC condition
requires $\m_b=0$ for all $T\leq T_c$.

The gap equation (\ref{thouless}) tells us that $t_\pg(Q)$ is highly
peaked around $Q=0$, so we can approximate $\S_\pg$ as
\begin{eqnarray}
\label{spg} \S_\pg(K)\simeq -\D_\pg^2\cg_0(-K),\;\;T\leq T_c,
\end{eqnarray}
where we have defined the pseudogap parameter via
\begin{eqnarray}
\label{pseudogap} \D^2_\pg=-\frac{1}{\b V}\sum_{Q} t_\pg(Q).
\end{eqnarray}
The total self-energy now is
\begin{eqnarray}
\label{sfull} \S(K)= -\D^2\cg_0(-K),
\end{eqnarray}
with $\D^2=\D^2_\sf+\D^2_\pg$. It is clear that $\D_\pg$ also
contributes to the energy gap in quasi-nucleon excitation.
Physically, the pseudogap $\D_\pg$ below $T_c$ can be interpreted as
an extra contribution to the excitation gap of nucleon
quasi-particle: an additional energy is needed to overcome the
residual attraction between nucleons in a thermal excited pair to
produce fermion-like quasi-particles. One should note that the
$\D_\pg$ is associated with the fluctuation of the pairs
$\D^2_\pg\sim\lan\D^2\ran-\lan\D\ran^2$~\cite{Chen:1998zz,prb61:11662},
hence it does not lead to superfluid (symmetry breaking).

With \eq{spg}, the pair susceptibility reads,
\begin{eqnarray}
\c(Q)&=&\frac{1}{V}\sum_\bk\Bigg[\frac{E_\bk+\x_
\bk}{2E_\bk}\frac{n_F(-\x_{\bq-\bk})-n_F(E_\bk)}{E_\bk+\x_{\bq-\bk}-q_0-i0^+}\non
&-&\frac{E_\bk-\x_
\bk}{2E_\bk}\frac{n_F(E_\bk)-n_F(\x_{\bq-\bk})}{E_\bk-\x_{\bq-\bk}+q_0+i0^+}\Bigg]\non
&=&\frac{1}{V}\sum_{\bk,s=\pm}\frac{sE_\bk+\x_\bk}{2sE_\bk}\frac{n_F(-sE_\bk)-n_F(\x_{\bq-\bk})}{sE_\bk+\x_{\bq-\bk}-q_0},\non
\end{eqnarray}
with $E_\bk=\sqrt{\xi_\bk^2+\D^2}$. The number equation remains
unchanged except the replacement of $\D_\mf\ra\D$.

Furthermore, the gap equation (\ref{thouless}) suggests that we can
make the following pole approximation to the pair propagator or $T$
matrix $t_\pg(Q)$ as
\begin{eqnarray}
\label{tpg1} t_\pg(Q)\simeq\frac{Z^{-1}}{q_0-\bq^2/(2m_b)},
\end{eqnarray}
where the residue $Z^{-1}$ and effective ``boson" mass are given by
\begin{eqnarray}
Z&=&\frac{\pt\c}{\pt q_0}\bigg|_{Q=0},\non
\frac{Z}{m_b}&=&-\frac{1}{3}\frac{\pt^2\c}{\pt \bq^2}\bigg|_{Q=0}.
\end{eqnarray}
We stress here that in general, the expansion of $t^{-1}_\pg(Q)$
should also contain a term $\propto q_0^2$, but such term does not
bring qualitative change to the crossover
physics~\cite{phyrept2005}, hence we neglect it in \eq{tpg1}.

A straightforward calculation gives
\begin{eqnarray}
\label{z}
Z&=&\frac{1}{V}\sum_\bk\sum_{s=\pm}\frac{s}{2E_\bk}\frac{n_F(E_\bk)-n_F(s\x_\bk)}{E_\bk-s\x_\bk}\non&=&\frac{1}{\D^2}\ls
\frac{\r}{4}-\frac{1}{V}\sum_\bk n_F(\x_\bk)\rs,
\end{eqnarray}
and
\begin{eqnarray}
\label{massboson}
\frac{Z}{m_b}&=&\frac{1}{V}\sum_\bk\sum_s\frac{1}{2sE_\bk}\bigg[\frac{1}{m}\frac{n_F(-sE_\bk)-n_F(\x_\bk)}{sE_\bk+\x_\bk}\non
&&-\frac{2\bk^2}{3m^2}\lb\frac{n_F(-sE_\bk)-n_F(\x_\bk)}{(sE_\bk+\x_\bk)^2}+\frac{n_F'(\x_\bk)}{sE_\bk+\x_\bk}\rb
\bigg]\non&=&\frac{Z}{m}-\frac{1}{V}\sum_\bk\frac{2\bk^2}{3m^2\D^2}n_F'(\x_\bk)-\frac{1}{V}\sum_\bk\frac{\bk^2}{3m^2E_\bk\D^4}\non
&&\!\!\!\!\!\!\!\times\left\{(E_\bk^2+\x_\bk^2)[1-2n_F(E_\bk)]-2E_\bk\x_\bk[1-2n_F(\x_\bk)]\right\}.\non
\end{eqnarray}
The expression in the square bracket of right-hand-side of \eq{z} is
nothing but one-half the density of the pairs $\r_b/2$, we then have
$\r_b=2Z\D^2$.

Substituting \eq{tpg1} into \eq{pseudogap} leads to
\begin{eqnarray}
\label{pseudogap1} \D^2_\pg&=&\frac{1}{ZV}\sum_\bq
n_B[\bq^2/(2m_b)]\non&=&\frac{1}{Z}\lb\frac{Tm_b}{2\p}\rb^{3/2}\z\lb3\over2\rb,
\end{eqnarray}
where $n_B(x)=1/[\exp{(\b x)}-1]$ is the Bose-Einstein function and
a vacuum term was regularized out. It should be stressed that at
zero temperature $\D^2_\pg=0$, hence the $G_0G$ scheme yields the
BCS ground state. One should note that $\D^2_\pg=\r_b^{\rm
uncondensed}/2Z$, and hence $\D^2_\sf=\r_b^{\rm condensed}/2Z$.

Now, \eq{thouless}, \eq{pseudogap1}, as well as number equation are
coupled to determine the total excitation gap $\D$, the pseudogap
$\D_\pg$ and the nucleon chemical potential $\m$ at given density
and temperature below $T_c$. In short, they are
\begin{eqnarray}
\label{gapdensity}
1&=&\frac{g}{V}\sum_\bk\frac{1}{2E_\bk}\ls1-2n_F(E_\bk)\rs,\non
\r&=& \frac{2}{V}\sum_\bk\ls1-\frac{\x_\bk}{E_\bk}\lb1-2
n_F(E_\bk)\rb\rs,\non
\D^2_\pg&=&\frac{1}{Z}\lb\frac{Tm_b}{2\p}\rb^{3/2}\z\lb3\over2\rb.
\end{eqnarray}

\subsection {$G_0G$ formalism above $T_c$}\label{above}
Above $T_c$, \eq{thouless} does not apply, hence \eq{spg} no longer
holds. To proceed, we extend our more precise $T\leq T_c$ equations
to $T>T_c$ in the simplest fashion. We will continue to use
\eq{sfull} to parameterize the self-energy but with $\D=\D_\pg$, and
ignore the finite lifetime effect associated with normal state
pairs. It was shown that this is still a good approximation when
temperature is not very high~\cite{phyc1999,phyrept2005,Chien:2009}.
The $T$ matrix $t_\pg(Q)$ at small $Q$ can be approximated now as
\begin{eqnarray}
\label{tpg2} t_\pg(Q)\simeq\frac{Z^{-1}}{q_0-\O_\bq},
\end{eqnarray}
where $\O_\bq=\bq^2/(2m_b)-\m_b$. Since there is no condensation in
normal state, the effective pair chemical potential $\m_b$ is no
longer zero, instead, it should be calculated from
\begin{eqnarray}
\label{gap2}
Z\m_b&\equiv&t^{-1}(0)=-\frac{1}{g}+\c(0)\non&=&-\frac{1}{g}+\frac{1}{V}\sum_\bk\frac{1-2n_F(E_\bk)}{2E_\bk}.
\end{eqnarray}
This is used as the modified gap equation. Similarly, above $T_c$
the pseudogap $\D_\pg$ is determined by
\begin{eqnarray}
\label{pseudogap2} \D^2_\pg&=&\frac{1}{ZV}\sum_\bq
n_B(\O_\bq)\non&=&\frac{1}{Z}\lb\frac{Tm_b}{2\p}\rb^{3/2}{\rm
Li}_{3\over2}\lb e^{\m_b/T}\rb,
\end{eqnarray}
where ${\rm Li}_n(z)$ is the polylogarithm function. Then \eq{gap2},
\eq{pseudogap2} and the number equation which remains unchanged
determine $\D_\pg$, $\m$ and $\m_b$.

In summary, at $T>T_c$, the order parameter is zero, and
$\D=\D_\pg$. The closed set of equations determining $\D$, $\m$ and
$\m_b$ is
\begin{eqnarray}
\label{gapdensity2}
Z\m_b&=&-\frac{1}{g}+\frac{1}{V}\sum_\bk\frac{1-2n_F(E_\bk)}{2E_\bk},\non
\r&=& \frac{2}{V}\sum_\bk\ls1-\frac{\x_\bk}{E_\bk}\lb1-2
n_F(E_\bk)\rb\rs,\non
\D^2_\pg&=&\frac{1}{Z}\lb\frac{Tm_b}{2\p}\rb^{3/2}{\rm
Li}_{3\over2}\lb e^{\m_b/T}\rb.
\end{eqnarray}

\subsection {Thermodynamics}\label{thermo}
The thermodynamics of the matter are governed by the thermodynamic
potential, which reads
\begin{eqnarray}
\label{o} \O&=&\O_f+\O_b,
\end{eqnarray}
where $\O_f$ and $\O_b$ are the contributions from nucleons and
thermal excited pairs,
\begin{eqnarray}
\label{of}
\O_f&=&2\D^2\c(0)-\frac{4T}{V}\sum_\bk\ls\frac{E_\bk-\xi_\bk}{2}+\ln{\lb1+e^{-\b E_\bk}\rb}\rs,\non\\
\label{ob} \O_b&=&\frac{2}{\b V}\sum_\bq\ln{\lb1-e^{-\b\O_\bq}\rb}.
\end{eqnarray}
Other thermodynamic quantities can be derived from $\O$, \eg, the
entropy density is given by $s=-\pt\O/\pt T$, and the specific heat
$c_V$ is given by $c_V=T\pt s/\pt T$.

\section {Numerical Results}\label{numerical}
We discuss now the results obtained by numerically solving
Eq.~(\ref{gapdensity}) for $T\leq T_c$ and Eq.~(\ref{gapdensity2})
for $T>T_c$. We will mainly focus on the intermediate (the crossover
region, see \fig{muxia}) and low density regions, since the high
density region is proven to be well understood in BCS theory. We
begin with the results concerning the critical temperature for
superfluid transition in the BCS-BEC crossover.

\subsection {BCS-BEC crossover and critical temperature}\label{crossover}
At zero temperature, the $G_0G$ formalism reproduces the usual BCS
theory. In order to have a quantitative examination of the BCS-BEC
crossover, it is convenient to define the condensed $np$ Cooper pair
wave function at zero temperature,
\begin{eqnarray}
\label{wavefunc} \j(\br)&\equiv&C\lan
BCS|a^\dag_{n\ua}(\bx)a^\dag_{p\da}(\bx+\br)|BCS\ran\non
&=&C'\idp{3}{\bk}\j(\bk)e^{i\bk\cdot\br},
\end{eqnarray}
where $a_{n\s}^\dag$ ($a_{p\s}^\dag$) is the creation operator of
neutron (proton) with spin $\s$ and $\j(\bk)$ are the anomalous
density distribution function
\begin{eqnarray}
\label{anomalousdensity} \j(\bk)&=&\lan
BCS|a^\dag_{n\ua}(\bk)a^\dag_{p\da}(-\bk)|BCS\ran\non&=&\frac{\D}{2E_\bk}.
\end{eqnarray}
After substituting \eq{anomalousdensity} into the number and gap
equations, we get the following Schr\"odinger-like equation,
\begin{eqnarray}
\label{schronp} \frac{\bk^2}{m}\j(\bk)-g(1-2n_\bk)\idp{3}{\bk'}
\j(\bk')=2\m\j(\bk).\non
\end{eqnarray}
In the limit of vanishing density, $n_\bk\ra0$, this equation goes
over into the Schr\"odinger equation for the $np$ bound states (the
deuterons) in the center-of-mass frame, and the chemical potential
$2\m$ then plays the role of the binding energy. Hence, one expects
that at sufficiently low density and low temperature, the symmetric
nuclear matter should be in the BEC phase.

To have more quantitative description of the BCS-BEC crossover, we
define other characteristic quantities: the mean-square-root size of
the $np$ pair,
\begin{eqnarray}
\label{xi} \x^2=\frac{\int d^3\bx \bx^2 |\j(\bx)|^2}{\int d^3\bx
|\j(\bx)|^2},
\end{eqnarray}
and the $s$-wave scattering length $a$ that relates the coupling
constant $g$ to the low-energy limit of the two-body $T$ matrix of
$np$ scattering in vacuum,
\begin{eqnarray}
\label{scatter} \frac{m}{4\p a}=-\frac{1}{g}+\int
\frac{d^3\bk}{(2\p)^3}\frac{m}{\bk^2}.
\end{eqnarray}
In BCS region, $\x$ is expected to be larger than the average
distance between neutron and proton $d\equiv(\r/2)^{-1/3}$ and at
the same time the scattering length $a$ should be negative to ensure
that the interaction between neutron and proton is attractive. In
the BEC region, however, $\x/d$ should be very small reflecting the
compactness of the pair, and the scattering length will be positive
to guarantee the appearance of two-body bound state.
\begin{figure}[!htb]
\begin{center}
\includegraphics[width=8.5cm]{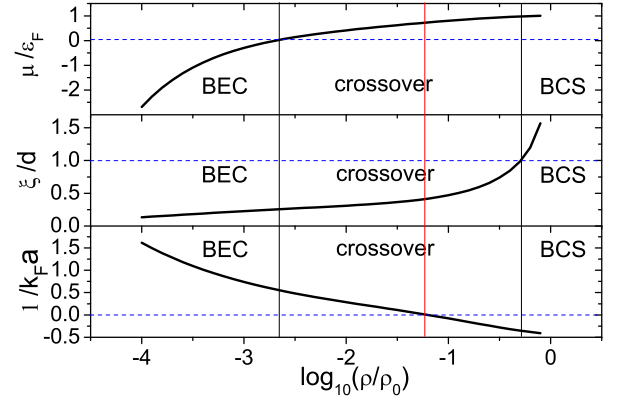}
\caption{(Color online) The nucleon chemical potential over Fermi
energy ratio $\m/\ve_F$, the scaled pair size $\x/d$, and the scaled
scattering length of $np$ collision $1/(k_F a)$ as functions of
density. The right and left vertical lines that separate the BCS,
BEC, and crossover regions are, respectively, determined by the
conditions $\x/d=1$, and $\m/\ve_F=0$. The red vertical line in
crossover region denotes the unitary point where $1/(k_F a)=0$.}
\label{muxia}
\end{center}
\end{figure}

In \fig{muxia}, we show the nucleon chemical potential over Fermi
energy ratio $\m/\ve_F$, the scaled mean-square-root size $\x/d$,
and the inverse scattering length $1/(k_F a)$ as functions of
nuclear density. Although $\x$ itself is not a monotonous function
of $\r$, the scaled one goes down monotonously with decreasing
density and finally approaches zero at zero density. The right
vertical line around $\r/\r_0\sim 0.5$ indicates the position of
$\x=d$ which can be used to separate the BCS (weak coupling) region
from the crossover (intermediately strong coupling) region. The
chemical potential roughly equals the Fermi energy at BCS region,
but it drops down with decreasing density and becomes negative below
$\r/\r_0\sim0.002$. The position where $\m$ changes sign can be
regarded as the boundary between BEC (strong coupling) region in
which $\m$ is negative and other region with positive $\m$. The
third panel shows that $1/(k_F a)$ increases with decreasing density
and becomes positive after $\r/\r_0\sim 0.06$. This turning point is
called the unitary limit, which we indicate by a red vertical line
in the figure. We will discuss unitary limit in next subsection.
\begin{figure}[!htb]
\begin{center}
\includegraphics[width=8cm]{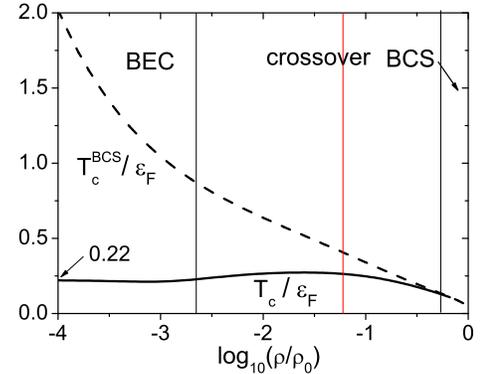}
\caption{(Color online) The critical temperature $T_c$ scaled by the
Fermi energy $\ve_F$ as a function of density. Also shown is the BCS
prediction (dashed line). } \label{tc}
\end{center}
\end{figure}

The numerical result for the critical temperature is shown in
\fig{tc}. The dashed line shows the critical temperature over Fermi
energy ratio given by BCS theory, which blows up quickly when
density goes down in the crossover and BEC regions ($T_c^\bcs$ is
almost ten times larger than $T_c$ at density $\r=0.0001\r_0$).
Physically, the ratio $T_c^{\rm BCS}/\ve_F$ as well as
$\D_{T=0}/\ve_F$ measures the strength of the attraction between
neutron and proton. However, due to the lack of pairing fluctuation
effect, in crossover and BEC regions the BCS theory does not give
correct critical temperature for superfluid/normal transition which
is mainly determined by the bosonic degree of freedom in these
regions. The solid line is for $T_c$ obtained from
Eq.~(\ref{gapdensity}). We can see that the evolution of $T_c$ is
smooth and the superfluid phase transition is second order in the
whole density region. Also, it can be seen that $T_c$ is not a
monotonous function of density: There is a local maximum in $T_c$
curve which is roughly located around the unitary point. One should
notice that a similar local maximum also appears in the famous
Nozieres$-$Schmitt-Rink approach for $T_c$~\cite{Nozieres:1985zz}.
At low density limit, all the nucleons participate into the
deuterons which are long lived at temperature lower than $T_c$, the
system is essentially a deuteron gas and the superfluid is totally
due to the BEC of deuterons. In this case, we have, at $T_c$, $2 Z
\D_\pg(T=T_c)=\r_b^{\rm uncondensed}=\r/2$. Solving out $T_c$, we
get,
\begin{eqnarray}
\label{tcbec} T_c&=&\frac{2\p}{m_b}\ls\frac{\r}{4\z(3/2)}\rs^{2/3}.
\end{eqnarray}
This is just the BEC transition temperature for boson of mass $m_b$.
Adopting that $m_b\approx2m$, we arrived at the well-known result,
\begin{eqnarray}
\label{tcbec2} T_c&\approx&0.218 \ve_F,
\end{eqnarray}
which coincides well with our numerical result.

It should be stressed that the BCS critical temperature $T_c^{\rm
BCS}$ was found to be a good approximation for the pair dissociation
temperature $T^*$(above which the pairs are essentially dissociated
by thermal motion of the participators)~\cite{phyc1999}. So it is a
pseudogap dominated region in between $T_c$ and $T_c^{\rm BCS}$.

\begin{figure}[!htb]
\begin{center}
\includegraphics[width=7.5cm]{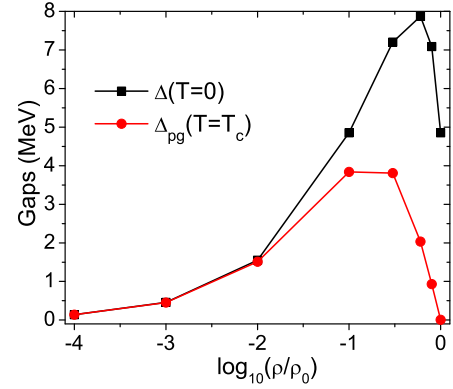}
\caption{(Color online) The excitation gap $\D$ at zero temperature
and the pseudogap at $T_c$ as functions of density. } \label{delt}
\end{center}
\end{figure}

Corresponding to the evolution of the critical temperature, it is
indicative to see how the pseudogap evolves. In \fig{delt}, we plot
the zero temperature excitation gap $\D(T=0)$ as well as the
pseudogap $\D_\pg(T=T_c)$ at $T_c$. It can be seen that at low
density $\r\lesssim0.01\r_0$, $\D_\pg(T=T_c)$ is roughly equal to
$\D(T=0)$ (but they do not completely coincide) reflecting the
strong coupled nature in this case; however, at high density,
$\D_\pg(T=T_c)$ is much smaller than $\D(T=0)$, indicating that the
pairing fluctuation is not essential there and BCS theory can work
well. In the following subsections, we will focus on the crossover
and BEC regions where BCS theory is not applicable at finite
temperature.

\subsection {Unitary matter}\label{unitary}
As shown in last subsection, the $np$ scattering length $k_F a$ is
infinite at the unitary point $\r\approx0.06\r_0$. We call the
nuclear matter at this point a unitary matter. It is interesting
because it exhibits universal behaviors, \ie, the physical
properties of unitary matter are independent of the details of the
interactions~\cite{Ho:2004}. Hence, the unitary nuclear matter
behaves just like unitary cold atomic Fermi gas which has been
realized in laboratory through the Feshbach resonance. For unitary
matter, the unique characteristic scale is given by the Fermi
momentum $k_F$, so we have $\m_{T=0}=\z\ve_F$, $\D_{T=0}=\g\ve_F$,
$T_c=\a\ve_F$, $\D_\pg(T=T_c)=\l\ve_F$, \etc, with
$\z,\,\g,\,\a,\,\l$, \etc, being universal constants. In our $G_0G$
scheme, the universal coefficients are given by
$\z\approx0.59,\,\g\approx0.64,\,\a\approx0.26$ and $\l\approx0.53$.
For comparison, we would like to list the values obtained by
Monte-Carlo techniques: $\z\approx0.42$~\cite{Carlson:2005kg},
$\g\approx0.50$~\cite{Carlson:2005kg},
$\a\approx0.157$~\cite{Burovski:2006zz} or
$0.25$~\cite{akkineni:2007}. Our results are larger than Monte-Carlo
values. It is easy to show that the energy per particle in unitary
matter is $E/N=\z (E/N)_{\rm free}$ where $(E/N)_{\rm free}$ is
energy per particle for free fermion gas. Moreover, the equation of
state of unitary matter is the same as free fermion gas, $\ve=3P/2$,
with $P$ being the pressure.
\begin{figure}[!htb]
\begin{center}
\includegraphics[width=7.5cm]{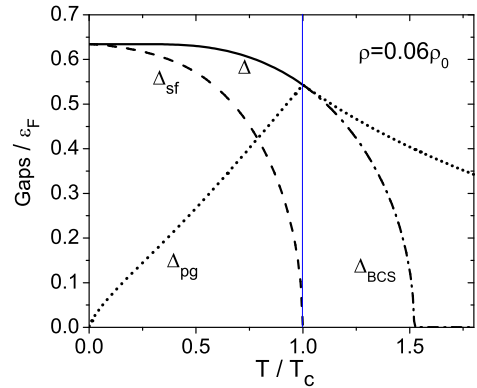}
\caption{(Color online) The superfluid gap $\D_\sf$ (dashed line),
pseudogap $\D_\pg$ (dotted line) and the total gap $\D$ (solid line)
as functions of the temperature at unitary point $\r=0.06\r_0$. The
BCS result (dot-dashed line) above $T_c$ is also shown.}
\label{gap1}
\end{center}
\end{figure}

In \fig{gap1}, we plot $\D_\sf$ (in units of $\ve_F$, the same
below),$\D_\pg$, and $\D$ as functions of temperature for unitary
matter. As a comparison, we also plot the BCS result $\D_{\rm BCS}$
above $T_c$. As we can see from the figure, with decreasing
temperature below $T_c$, $\D_\pg(T)$ is a monotonically decreasing
function from its maximum value at $T_c$ and it essentially vanishes
at $T=0$ roughly according to $\D_\pg(T)\propto T^{3/4}$ [see
\eq{pseudogap1}], while $\D_\sf(T)$ and $\D(T)$ both increase
monotonically and become coincident at $T=0$. Such kinds of
temperature dependence reflect the fact that the pseudogap is due to
the thermally excited pairs: When $T$ grows higher and higher, more
and more pairs are excited from the condensate and at $T_c$ all the
condensed pairs are thermally excited; after that the thermal motion
of nucleons begins to dissociate the pairs and hence $\D_\pg$ (more
exactly, $Z\D_\pg^2$) begins decreasing above $T_c$. In addition,
the critical temperature $T_c$ is smaller than the BCS prediction
which shows the fact that the pairing fluctuation tends to destroy
the order of the system. Although the physical picture is clear, our
formalism cannot be applied to very high temperature, where the
effects of finite life-time of the pairs become significant which
are not included in our formalism.
\begin{figure}[!htb]
\begin{center}
\includegraphics[width=7.5cm]{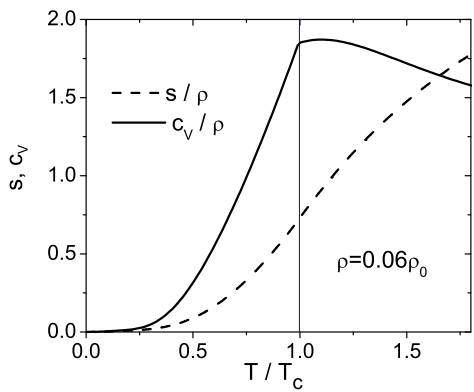}
\caption{(Color online) The entropy density and the specific heat of
unitary matter as functions of temperature. Unlike the weak-coupling
BCS case, there is no jump for the specific heat at $T_c$.}
\label{cv1}
\end{center}
\end{figure}

One should note that $\D$ and its derivative $d\D/d T$ are
continuous at $T_c$. This is very different from the BCS case, where
$d\D_\bcs/d T$ is discontinuous at $T_c^\bcs$. Such difference could
be reflected in thermodynamic quantities such as the specific heat.
In \fig{cv1}, we illustrate the entropy density and the specific
heat $c_V$ for unitary matter. We compute $c_V$ through $c_V=T\pt
s/\pt T$ which involves the derivatives $\pt \m/\pt T$, $\pt\D/\pt
T$, $\pt m_b/\pt T$, and $\pt \m_b/\pt T$, so it is a nontrivial
calculation. As is well-known, for the weak-coupling BCS case, the
specific heat has a jump $\D c_V\propto d\D^2/d T$ at $T_c^\bcs$
which reflects the sudden opening of the excitation gap. For unitary
matter, however, we found a continuous $c_V$ at $T_c$. This
continuity of $c_V$ reflects the previous existence of the
excitation gap above $T_c$ due to the pairing fluctuation. It may
serve as an experimentally accessible signal for the existence of
the pseudogap in the normal phase.

\subsection {Deuteron gas}\label{deuteron}
When density is very low, say $\r<0.002\r_0$ from \fig{muxia}, the
symmetric nuclear matter is effectively a Bose gas of deuteron. We
in this subsection study the properties of deuteron gas based on our
$G_0G$ formalism. First, we observed from \fig{muxia} and \fig{tc}
that when $\r\ra0$, $-\m\gg T_c$. In this case, $\D$, $\m$ and $Z$
are almost temperature independent below $T_c$ reflecting the strong
$np$ attraction. Then for $T<T_c$, the governing equations become
(expanding in powers of $a^3 \r$ and
$\D^2/\m^2$)~\cite{phyrept2005},
\begin{eqnarray}
\label{eq:bec} \frac{m}{4\p
a}&=&\frac{1}{V}\sum_\bk\lb\frac{m}{\bk^2}-\frac{1}{2E_\bk}\rb\approx\frac{m\sqrt{2m|\m|}}{4\p}\lb1+\frac{\D^2}{16\m^2}\rb,\non
\r&=&
\frac{2}{V}\sum_\bk\lb1-\frac{\x_\bk}{E_\bk}\rb\approx\frac{m^2\D^2}{2\p\sqrt{2m|\m|}},\non
\D^2_\pg&\approx&\frac{4\D^2}{\r}\lb\frac{T
m}{\p}\rb^{3/2}\z\lb3\over2\rb.
\end{eqnarray}
Hence at low temperature, we have
\begin{eqnarray}
\label{eq:bec2} \D^2&\approx&\frac{2\p \r}{m^2 a}\lb1-\frac{\p a^3
\r}{2}\rb,\non \D_\pg^2&\approx&\frac{8\p}{m^2
a}\z\lb\frac{3}{2}\rb\lb\frac{m T}{\p}\rb^{3/2}\lb1-\frac{\p a^3
\r}{8}\rb,\non \m&\approx&-\frac{1}{2m a^2}\lb1-\p a^3 \r\rb.
\end{eqnarray}
These relations give how $\D(\r)$, $\D_\pg(\r)$, $\m(\r)$, and
$m_b(\r)$ evolve with $\r$ at $T<T_c$ in the deep BEC region.

\begin{figure}[!htb]
\begin{center}
\includegraphics[width=7.5cm]{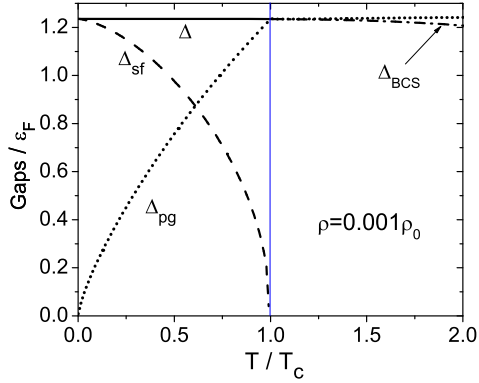}
\caption{(Color online) The superfluid gap $\D_\sf$, pseudogap
$\D_\pg$ and the total gap $\D$ as functions of the temperature at
density $\r=0.001\r_0$. The BCS result is also shown above $T_C$.}
\label{gap3}
\end{center}
\end{figure}

Next, let us study the temperature dependence of these
characteristic quantities. To specify the problem, we fix the
density as $\r=0.001\r_0$. By solving the coupled equations
(\ref{gapdensity}), we get the transition temperature
$T_c\simeq0.22\,\ve_F\approx0.08$ MeV. In \fig{gap3}, we show the
superfluid gap $\D_\sf$ , the pseudogap $\D_\pg$, the total
excitation gap $\D$ as functions of temperature. Due to the stronger
attraction, unlike for the unitary matter, now $\D(T)$ is almost a
constant below $T_c$, and at $T=0$, $\D/\ve_F$ is even larger than
1. But near zero temperature, $\D_\pg(T)$ still behaves as
$\D_\pg\propto T^{3/4}$ as shown in Eq.~(\ref{eq:bec2}).

\begin{figure}[!htb]
\begin{center}
\includegraphics[width=7.5cm]{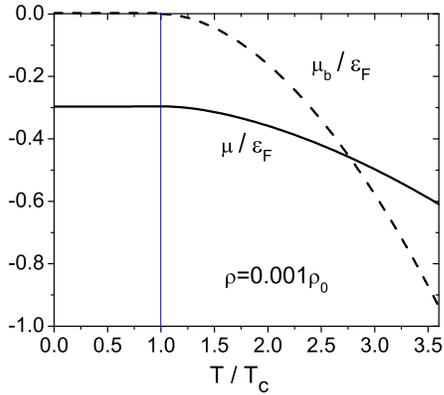}
\caption{(Color online) The nucleon chemical potential and the
effective deuteron chemical potential as functions of $T$ at density
$\r=0.001\r_0$.} \label{mu3}
\end{center}
\end{figure}

\begin{figure}[!htb]
\begin{center}
\includegraphics[width=7.5cm]{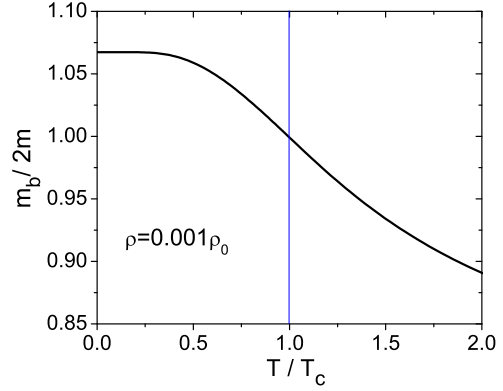}
\caption{(Color online) The effective deuteron mass parameter as a
function of $T$ at density $\r=0.001\r_0$.} \label{mb3}
\end{center}
\end{figure}

In \fig{mu3}, we give the temperature dependence of nucleon chemical
potential $\m(T)$ and effective deuteron chemical potential
$\m_b(T)$. Below $T_c$, $\m_b$ is zero meaning that BEC superfluid
is formed. Above $T_c$, both $\m$ and $\m_b$ decrease, corresponding
to the thermal dissociating effect. Just above $T_c$, simple
calculation leads to that $\m(T)-\m(T_c)\propto\m_b\propto
(T-T_c)^2$. Since $\m_b$ is related to $\m$ and the deuteron binding
energy $E_b$ through $\m_b=2\m+E_b$, we get that the binding energy
$E_b$ at zero temperature is roughly $0.6\,\ve_F$ at $\r=0.001\r_0$
from \fig{mu3}. The effective deuteron mass $m_b(T)$ is shown in
\fig{mb3}. It is seen that at low temperature $m_b$ is almost a
constant, but it drops when temperature becomes higher. We note here
that $m_b$ can be regarded as the medium renormalized deuteron mass
only in deep BEC region~\cite{Haussmann,Fetter}. It is renormalized
because it contains indirectly the deuteron-deuteron interaction
through the nucleon-deuteron coupling in the nucleon self-energy.
Hence, this effective mass is not equal to $2m-E_b$, as one may
intuitively expect. It is actually a parameter measuring the
effective size of the non-condensed pairs, hence it is also defined
in intermediate coupling and even weak coupling regions. The drop of
$m_b$ at high temperature simply indicates that the effective size
of the non-condensed pair is enlarged by the thermal motion of
participate nucleons.

\begin{figure}[!htb]
\begin{center}
\includegraphics[width=7.5cm]{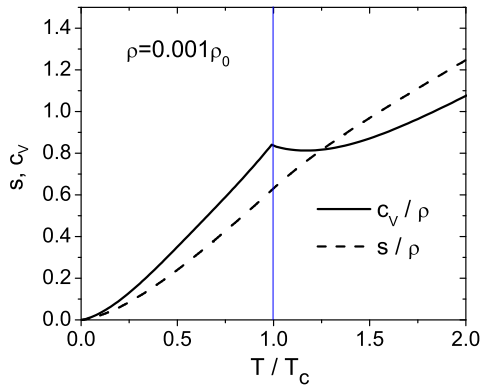}
\caption{(Color online) The entropy density and the specific heat of
symmetric nuclear matter as functions of temperature at density
$\r=0.001\r_0$.} \label{cv3}
\end{center}
\end{figure}

Finally, we depict the entropy density and the specific heat for
deuteron gas in \fig{cv3}. One should note that at low temperature,
the behavior of $c_V$ is very different from the prediction of BCS
theory: It shows $T^{3/2}$ dependence at low $T$ rather than an
exponential suppression. Actually, since the condensate does not
contribute to entropy, $c_V$ at $T\ll T_c$ contains contributions
from quasi-nucleons and from thermal excited pairs. The former
contribution is just the BCS theory result,
\begin{eqnarray}
c_V^\bcs\propto \lb\frac{\D_0}{T}\rb^{1/2}e^{-\D_0/T},\; T\ll T_c,
\end{eqnarray}
with $\D_0=\D_{T=0}$. The latter one is dominated by
$T\pt\D_\pg^2/\pt T$ for other quantities are almost independent of
$T$ [see \eq{eq:bec2}], hence
\begin{eqnarray}
c_V^\pg\propto T^{3/2},\; T\ll T_c,
\end{eqnarray}
which dominates $c_V$ at low temperature. At the phase transition
point, similarly with unitary matter, due to the continuity of the
temperature derivative of excitation gap, $c_V$ does not get a
discontinuity, but a $\l$-type behavior. Now, both the low
temperature and the critical behaviors of $c_V$ are quite similar
with the situation found in ideal BEC superfluid. It indicates that
the symmetric nuclear matter at very low density is a nearly ideal
deuteron gas.

\begin{figure}[!htb]
\begin{center}
\includegraphics[width=7.5cm]{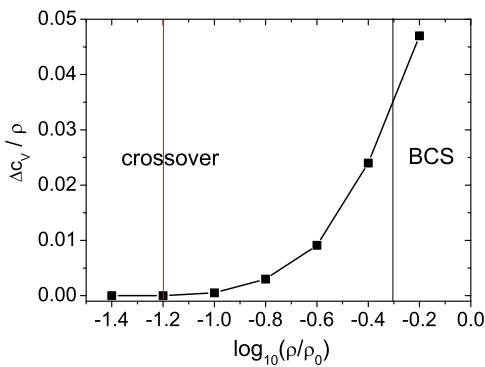}
\caption{(Color online) The specific heat jump at $T_c$ over density
ratio $\D c_V/\r$ as a function of density.} \label{deltcv}
\end{center}
\end{figure}

Figures.\ref{cv1} and \ref{cv3} inspire us that the specific heat
jump at $T_c$ may serve as a possible thermodynamic signal for
BCS-BEC crossover. We hence draw in \fig{deltcv} the specific heat
jump $\D c_V\equiv c_V(T_c-0^+)-c_V(T_c+0^+)$ at $T_c$ over density
ratio as a function of density. As density decreases, $\D c_V/\r$
monotonously decreases and vanishes when density is smaller than
$0.06\r_0$ which is just the unitary point. The physical reason for
such kind of behavior is clear: as density decreases the system
becomes more and more bosonic and the finite jump of the specific
heat at $T_c$ which is a typical BCS feature gets suppressed.

\section {Summary and Discussion}\label{summary}
As is well known, BCS theory is only applicable to weak coupling
system or at zero temperature since it does not contain any pairing
fluctuation effects. To study the BCS-BEC crossover at finite
temperature, it is necessary to go beyond BCS description. We, in
this article, studied the effects of pseudogap due to pairing
fluctuation on the BCS-BEC crossover problem in symmetric nuclear
matter. For this purpose, we adopted a $T$-matrix method based on a
$G_0G$ approximation for the pair susceptibility. This method is a
natural extension of BCS theory and has been widely used in
theoretical studies of high $T_c$ superconductor and BCS-BEC
crossover problems in cold fermion atoms.

The pseudogap is determined by the density of thermally excited $np$
pairs and vanishes at zero temperature. We found that its effects
are substantial for intermediate and strong coupling regions
(corresponding to intermediate and low-density regions) in the
BCS-BEC crossover when temperature is not zero. At high-density
region, the pseudogap is essentially small, and the $G_0G$ theory
recovers the BCS theory. Taking into account the pseudogap effects,
we calculated the critical temperature for superfluid phase
transition shown in \fig{tc}. At high density, $T_c$ follows the BCS
result, but at intermediate and low densities, it deviates from the
BCS prediction remarkably. At dilute limit, $T_c$ coincides with the
BEC transition temperature of dilute deuterons.

The pseudogap persists in both $T>T_c$ and $T<T_c$ regions. We
investigated how the pseudogap affects the properties of unitary
matter and dilute deuteron matter. At intermediate and low
densities, the significant result is that due to the pseudogap, the
specific heat is continuous at $T_c$. At low density, $c_V\propto
T^{3/2}$ at low temperature and has a continuous $\l$-type behavior
at $T_c$ just like an ideal BEC superfluid. The qualitative change
of the temperature dependence of specific heat from high density to
low density also indicates the BCS-BEC crossover. Moreover, as
density decreases, the jump of specific heat at $T_c$ decreases and
eventually disappears when $\r\lesssim0.06\r_0$ as shown in
\fig{deltcv}. This may serve as a thermodynamic signal for BCS-BEC
crossover.

We stress that the $G_0G$ approximation is, in principle, not
applicable at temperature much higher than $T_c$, since then the
thermal dissociation effect will result in a finite width of the
pairs which we did not take into account. Besides, symmetric nuclear
matter may favor the formation of $\a$ cluster~\cite{Ropke:1998qs},
but we did not consider this situation in present article. Finally,
since most of the nuclear systems in nature do not contain equal
numbers of neutrons and protons, it will be significant to extend
present formalism to asymmetric nuclear matter~\cite{Mao:2008wz}. We
leave all these challenges for future studies.

{\bf Acknowledgments:} We gratefully acknowledge H. Abuki, T.
Brauner and L. He for their careful reading of the manuscript and
numerous suggestions. We also thank D. Rischke, A. Sedrakian and P.
Zhuang for helpful discussions. This work is supported, in part, by
the Helmholtz Alliance Program of the Helmholtz Association,
Contract No. HA216/EMMI ``Extremes of Density and Temperature:
Cosmic Matter in the Laboratory'' and the Helmholtz International
Center for FAIR within the framework of the LOEWE (Landesoffensive
zur Entwicklung Wissenschaftlich-\"Okonomischer Exzellenz) program
launched by the state of Hesse.

\end{document}